\pdfoutput=1
\documentclass[twocolumn,prapplied,superscriptaddress,longbibliography]{revtex4-1}
\usepackage{graphicx,amsmath,amssymb}
\usepackage[colorlinks=true, citecolor=blue, urlcolor=blue, linkcolor=red]{hyperref}

\begin{document}
\title{Threshold for a Discrete-Variable Sensor of Quantum Reservoirs}
\author{Wei Wu}
\affiliation{Lanzhou Center for Theoretical Physics and Key Laboratory of Theoretical Physics of Gansu Province, Lanzhou University, Lanzhou, Gansu, China}
\author{Zhen Peng}
\affiliation{Lanzhou Center for Theoretical Physics and Key Laboratory of Theoretical Physics of Gansu Province, Lanzhou University, Lanzhou, Gansu, China}
\author{Si-Yuan Bai}
\affiliation{Lanzhou Center for Theoretical Physics and Key Laboratory of Theoretical Physics of Gansu Province, Lanzhou University, Lanzhou, Gansu, China}
\author{Jun-Hong An}
\email{anjhong@lzu.edu.cn}
\affiliation{Lanzhou Center for Theoretical Physics and Key Laboratory of Theoretical Physics of Gansu Province, Lanzhou University, Lanzhou, Gansu, China}

\begin{abstract}
Quantum sensing employs quantum resources of a sensor to attain a smaller estimation error of physical quantities than the limit constrained by classical physics. To measure a quantum reservoir, which is significant in decoherence control, a nonunitary-encoding sensing scheme becomes necessary. However, previous studies showed that the reservoir-induced degradation to quantum resources of the sensor makes the errors divergent with the increase of encoding time. We here propose a scheme to use $N$ two-level systems as the sensor to measure a quantum reservoir. A threshold, above which the shot-noise-limited sensing error saturates or even persistently decreases with the encoding time, is uncovered. Our analysis reveals that it is due to the formation of a bound state of the total sensor-reservoir system. Solving the outstanding error-divergency problem in previous studies, our result supplies an insightful guideline in realizing a sensitive measurement of quantum reservoirs.
\end{abstract}

\maketitle

\section{Introduction}
Quantum sensing aims at achieving a highly precise measurement to physical quantities with the help of quantum resources of a sensor~\cite{RevModPhys.89.035002}. The sensing error governed by classical physics is constrained by the so-called shot-noise limit (SNL)~\cite{PhysRevD.23.1693}. It has been demonstrated that the SNL can be beaten by using quantum entanglement~\cite{RevModPhys.90.035005,PhysRevLett.121.160502,Nagata726} and squeezing~\cite{PhysRevLett.122.173601,PhysRevLett.124.060402,PhysRevLett.113.103004,PhysRevLett.119.193601}. Many fantastic applications of quantum sensing have been made in gravitational-wave detection~\cite{PhysRevLett.123.231107,PhysRevLett.123.231108}, quantum radar~\cite{PhysRevLett.124.200503}, atom clocks~\cite{PhysRevLett.112.190403,Mehlst_ubler_2018}, magnetometers~\cite{Potts2019fundamentallimits,PhysRevX.10.011018,PhysRevLett.120.260503,PhysRevA.99.062330}, and thermometries~\cite{PhysRevA.98.050101,PhysRevB.98.045101}. A common feature of these applications is that the quantities of interest are encoded into the sensor state via a unitary dynamics~\cite{PhysRevLett.79.3865,PhysRevD.23.1693,Wang2019,Chabuda2020,PhysRevLett.122.090503,PhysRevResearch.1.032024,PhysRevApplied.13.024037,PhysRevX.9.041023,PhysRevLett.124.010507,PhysRevA.99.033807}. Such unitary-encoding scheme is applicable only in measuring the quantities of classical systems. When the ones of a quantum system are measured, the sensor-system coupling for quantity encoding inevitably results in a nonunitary dynamics of the sensor. How do we generalize the well-developed unitary-encoding sensing scheme to the nonunitary case and achieve a highly precise sensing of quantum systems in the nonunitary-encoding setting?

Recently, much attention has been focused on precisely measuring quantum reservoirs~\cite{PhysRevApplied.13.034045,PhysRevLett.123.230801,PhysRevLett.118.100401,PhysRevA.100.032108,PhysRevB.101.104306,PhysRevA.97.012126,PhysRevA.97.012125,Tamascelli_2020,SALARISEHDARAN2019126006,681442767}. The coupling of the reservoir to any microscopic system would cause the system to lose its quantum coherence, which is called decoherence. Controlling the detrimental impacts of decoherence on the relevant system is crucial in realizing quantum information processing and many other quantum technologies~\cite{LI20181,RevModPhys.88.021002,RevModPhys.89.015001,Rivas_2014,RevModPhys.89.041003,LI20181}.  Decoherence is essentially determined by the spectral density of the reservoir, which characterizes the system-reservoir coupling strength per unit frequency of the reservoir. Therefore, the grasping of the feature of the spectral density is a prerequisite for decoherence control \cite{PhysRevLett.87.270405,Soare2014}. However, in many occasions, the spectral density cannot be microscopically derived from first principle. Thus, a precise sensing to the spectral density of quantum reservoir is strongly necessary~\cite{PhysRevApplied.13.034045,PhysRevLett.123.230801,PhysRevLett.118.100401,PhysRevA.100.032108,PhysRevB.101.104306}. Several nonunitary-encoding schemes of sensing the quantum reservoir have been proposed. Unfortunately, as shown in previous works~\cite{PhysRevA.97.012126,PhysRevA.97.012125,Tamascelli_2020,SALARISEHDARAN2019126006,681442767}, although the nonunitary dynamics can successfully encode the spectral density into the sensor state, it meanwhile erases the quantum resource attached to the sensor, which causes the sensing error to become larger and larger with the increase of encoding time. Thus, solving the error-divergency problem in the long-encoding-time regime still remains as an open question and is necessary for any scheme to sense a quantum reservoir.

In this work, using $N$ two-level systems (TLSs) as the sensor, we propose a nonunitary-encoding sensing scheme to estimate the spectral density of a quantum reservoir. Going beyond the widely used pure-dephasing coupling and the Born-Markovian approximation, we exactly study the performance of an initial product and a Greenberger-Horne-Zeilinger (GHZ) entangled state of the sensor on the estimation error via the dissipative interactions for encoding. A threshold, above which the sensing error decreases with the encoding time as $t^{-1}$ for the initial product state and saturates to a finite value with the encoding time for the entangled state, is found. This is in sharp contrast to the error-divergency result in the previous works. Further study reveals that it is due to the formation of a bound state of the composite system consisting of each TLS and the reservoir. Supplying an efficient way to eliminate the longstanding error-divergency problem in sensing a quantum reservoir by manipulating the formation of the bound state, our scheme can be used to realize a SNL-type estimation to the spectral density of the quantum reservoir.

\section{Quantum parameter estimation}~\label{sec:sec2}
A quantum sensing scheme generally contains the steps of initial-state preparation, parameter encoding, and measurements. To sense a quantity $\theta$ of certain physical system, one first prepares a well-tailored quantum sensor in an initial state $\rho_\text{in}$. The sensed quantity $\theta$ is encoded into the sensor state $\rho_\theta$ via coupling the sensor to the physical system. Finally, one measures a chosen physical observable $\hat{O}$ of the sensor and infers the value of $\theta$ from the measurement results. In any quantum sensing process, one cannot completely eliminate the errors and estimate the quantity of interest precisely. According to quantum parameter-estimation theory \cite{Liu_2019}, whatever the observable $\hat{O}$ is measured, the ultimate estimation error of $\theta$ is constrained by the famous quantum Cram\'{e}r-Rao bound $\delta \theta\geq(\upsilon \mathcal{F}_{\theta})^{-1/2}$. Here $\delta\theta$ is the root mean square as the error of $\theta$, $\upsilon$ is the number of repeated measurements which is set to 1 in this work for explicitness, and $\mathcal{F}_{\theta}\equiv\mathrm{Tr}(\hat{L}^2_{\theta}\rho_\theta)$, with $\hat{L}_{\theta}$ being determined by $\partial_{\theta}\rho_\theta=(\hat{L}_{\theta}\rho_\theta+\rho_\theta\hat{L}_{\theta})/2$, is quantum Fisher information (QFI) describing the most information of $\theta$ contained in $\rho_\theta$. It can be readily found that the minimal sensing error is completely determined by the QFI: a larger QFI always means a smaller estimation error. Maximizing the value of QFI by optimizing the quantum resource in $\rho_\text{in}$ and the sensor-system interaction is the major objective of quantum sensing. When $\delta\theta\propto N^{-1/2}$, or equivalently, $\mathcal{F}_{\theta}\propto N$, with $N$ being the total particle number in the sensor, such scaling relation is called SNL. It has been demonstrated that the SNL can be surpassed by using entanglement \cite{Giovannetti1330,PhysRevLett.96.010401,PhysRevA.54.R4649,PhysRevLett.79.3865,Daryanoosh2018}, by the encoding process via quantum criticality~\cite{PhysRevLett.121.020402,PhysRevX.8.021022}, and by using nonlinear interactions \cite{Chen2018}.

\section{Quantum sensing of a dissipative reservoir}~\label{sec:sec3}
In order to precisely sense a quantum reservoir, we employ $N$ identical TLSs as the sensor. The information of the reservoir is encoded into the sensor via the individual interactions of each TLS with the reservoir described by the following Hamiltonian ($\hbar=1$):
\begin{equation}~\label{Ham}
\hat{H}=\sum_{l=1}^N\big\{\omega_{0}\hat{\sigma}_{l}^{+}\hat{\sigma}_{l}^{-}+\sum_{k}\big{[}\omega_{k}\hat{b}_{l k}^{\dagger}\hat{b}_{l k}+ (g_{k}\hat{\sigma}_{l}^{+}\hat{b}_{lk}+\text{H.c.})\big{]}\big\},
\end{equation}
where $\hat{\sigma}_{l}^{\pm}$ is the ladder operators between the ground state $|g\rangle$ and excited state $|e\rangle$ of the $l$th TLS with frequency $\omega_{0}$, $\hat{b}_{lk}$ denotes the annihilation operator of the $k$th mode with frequency $\omega_{k}$ of the reservoir coupled to the $l$th TLS, and $g_{k}$ denote their coupling strengths. The specific properties of the reservoir is described by its spectral density $J(\omega)\equiv\sum_{k}g_{k}^{2}\delta(\omega-\omega_{k})$, which in turn determines the decoherence effect caused by this reservoir to any quantum system.

Assuming the sensor-reservoir coupling is switched on at the initial time, thus we have the initial state of the total system as $\rho(0)\otimes\rho_{\mathrm{R}}(0)$ with $\rho(0)$ and $\rho_{\mathrm{R}}(0)=\bigotimes_{lk}|0_{lk}\rangle\langle 0_{lk}|$ being the initial sate of the sensor and the reservoirs, respectively. After tracing out the degrees of freedom of the reservoirs, we find that the dynamics of the sensor is governed by the non-Markovian master equation \cite{Wang_2017}
\begin{equation}~\label{masteq}
\dot{\rho}(t)=\sum_{l=1}^{N}\Big{\{}\frac{i}{2}\Omega(t)\big{[}\rho(t),\hat{\sigma}_{l}^{+}\hat{\sigma}_{l}^{-}\big{]}+\frac{1}{2}\gamma(t)\mathcal{\hat{L}}_{l}\rho(t)\Big{\}},
\end{equation}
where $\mathcal{\hat{L}}_{l}\rho(t)\equiv 2\hat{\sigma}_{l}^{-}\rho(t)\hat{\sigma}_{l}^{+}-\{\rho(t),\hat{\sigma}_{l}^{+}\hat{\sigma}_{l}^{-}\}$ describes the dissipation induced by reservoirs, $\Omega(t)\equiv-2\mathrm{Im}[\dot{c}(t)/c(t)]$ is the renormalized frequency of the $l$th TLS and $\gamma(t)\equiv-2\mathrm{Re}[\dot{c}(t)/c(t)]$ denotes the dissipation rate. Here, $c(t)$ is determined by the integro-differential equation
\begin{equation}~\label{ct}
\dot{c}(t)+i\omega_{0}c(t)+\int_{0}^{t}\nu(t-\tau)c(\tau)d\tau=0,
\end{equation}
with $c(0)=1$ and $\nu(x)\equiv\int_{0}^{\infty}J(\omega)e^{-i\omega x}d\omega$. It can be seen that the information of the reservoir characterized by the spectral density $J(\omega)$ has been successfully encoded in the sensor state via the master equation \eqref{masteq}.

It should be emphasized that, different from the widely used unitary evolution, the parameter encoding governed by Eq. \eqref{masteq} is a nonunitary dynamics of the sensor. Such nonunitary dynamics in turn causes decoherence to the sensor, which destroys the quantum resource carried by the sensor. Therefore, balancing the necessary parameter-encoding process and the decoherence effect in the nonunitary dynamics is crucial in achieving the high precision of our sensing scheme. Taking the Ohmic-family spectral density $J(\omega)=\eta\omega^{s}\omega_{c}^{1-s} e^{-\omega/\omega_{c}}$ as an example, we explore the performance of our sensing scheme to the reservoir. Here $\eta$ is a dimensionless coupling constant, $\omega_{c}$ is a cutoff frequency, and $s$ is the so-called Ohmicity parameter. Depending on the value of $s$, a quantum reservoir can be classified into three categories: a sub-Ohmic reservoir ($0<s<1$), an Ohmic reservoir ($s=1$), and a super-Ohmic reservoir ($s>1$) \cite{RevModPhys.59.1}. In this paper, $s$, $\omega_{c}$, and $\eta$ are the quantities to be estimated.

We consider two different initial states: an uncorrelated state $|\psi_{\mathrm{in}}\rangle=[(|e\rangle+|g\rangle)/\sqrt{2}]^{\otimes N}$ and a GHZ-type maximally entangled state $|\psi_{\mathrm{in}}\rangle=(|e\rangle^{\otimes N}+|g\rangle^{\otimes N})/\sqrt{2}$. The solution of the master equation \eqref{masteq} can be expressed in the Kraus representation as $\rho(t)=\hat{\Lambda}^{\otimes N}\rho(0)$ with $\hat{\Lambda}\rho(0)\equiv\sum_{j=1}^{2}\hat{K}_{j}\rho(0)\hat{K}_{j}^{\dagger}$, where $\hat{K}_{1}(t)=c(t)|e\rangle\langle e|+|g\rangle\langle g|$ and $\hat{K}_{2}(t)=(1-p_t)^{1/2}|g\rangle\langle e|$ with $p_t=|c(t)|^2$ \cite{Wang_2017,PhysRevA.84.022302,PhysRevA.87.032102}. We can derive the reduced density matrix of the sensor for the two initial states as
\begin{eqnarray}
\rho_{\mathrm{unc}}(t)&=&\big{\{}{p_t/ 2}|e\rangle\langle e|+[1-{p_t/ 2}]|g\rangle\langle g|\nonumber\\
&&+[{c(t)/ 2}|e\rangle\langle g|+\mathrm{H}.\mathrm{c}.]\big{\}}^{\otimes N},\label{singlequbit}\\
\rho_{\mathrm{GHZ}}(t)&=&\frac{1}{2}\big\{\big{[}p_t|e\rangle\langle e|+(1-p_t)|g\rangle\langle g|\big{]}^{\otimes N}+|g\rangle\langle g|^{\otimes N}\nonumber\\
  &&+[c(t)^{N}|e\rangle\langle g|^{\otimes N}+\mathrm{H}.\mathrm{c}.]\big\}.\label{nrhot}
\end{eqnarray}
After some algebra (see Appendix \ref{dervat}), we obtain the QFI of $\rho_{\mathrm{unc}}(t)$ as
\begin{eqnarray}
\mathcal{F}_\theta^\text{unc}(t)&=&N\bigg[|{\bf r}'(t)|^2+{[{\bf r}(t)\cdot{{\bf r}'(t)}]^2\over 1-|{\bf r}(t)|^2}\bigg],\label{uncff}
\end{eqnarray}
where $\mathbf{r}(t)=\big{(}\mathrm{Re}[c(t)],-\mathrm{Im}[c(t)],p_{t}-1\big{)}^{\mathrm{T}}$ being the Bloch vector and ${\bf r}'(t)=\partial_\theta{\bf r}(t)$. The one of $\rho_{\mathrm{GHZ}}(t)$ is $\mathcal{F}_\theta^\text{GHZ}(t)=\mathcal{F}_{\theta}^{(1)}+\mathcal{F}_{\theta}^{(2)}$ with (see Appendix \ref{dervat})
\begin{eqnarray}
\mathcal{F}_{\theta}^{(1)}=\sum_{i, j}\bigg{[}\lambda^{-1}_{i}{\lambda}_{i}^{\prime2}+4\lambda_{i}\langle\lambda'_{i}|\lambda'_{i}\rangle-
\frac{8\lambda_{i}\lambda_{j}}{\lambda_{i}+\lambda_{j}}|\langle\lambda_{i}|\lambda'_{j}\rangle|^{2}\bigg{]},~~~\label{ghzf1}\\
\label{ghzf2}
  \mathcal{F}_{\theta}^{(2)}=-\frac{Np_t^{\prime2}}{2p_t}\bigg\{\frac{Np_t(1-p_t)^{N}-(1-p_t)}{(1-p_t)^{2}}+Np_t^{N-1}\bigg{\}}.~~~
\end{eqnarray}
where $\lambda_i'=\partial_{\theta}\lambda_i$, $|\lambda_i'\rangle=\partial_\theta|\lambda_i\rangle$, and $\frac{1}{2}\{p_{t}^{N}|e\rangle\langle e|^{\otimes N}+[1+(1-p_{t})^{N}]|g\rangle\langle g|^{\otimes N}+(c_{t}^{N}|e\rangle\langle g|^{\otimes N}+\mathrm{H}.\mathrm{c}.)\}=\sum_i\lambda_i|\lambda_i\rangle\langle\lambda_i|$.

In the special case when the sensor-reservoir coupling is weak and the characteristic time scale of the reservoir is smaller than that of the sensor, we can safely apply the Markovian approximation to Eq.~(\ref{ct}). Under this approximation, we have
$c(t)\simeq \exp\{-\kappa t-i[\omega_{0}+\Delta(\omega_{0})]t\}$~\cite{PhysRevE.90.022122}, where $\kappa=\pi J(\omega_{0})$ determines the decay rate and $\Delta(\omega_{0})=\mathcal{P}\int_{0}^{\infty}\frac{J(\omega)}{\omega_{0}-\omega}d\omega$ is a frequency shift induced by the reservoirs. Substituting the above approximate $c(t)$ into Eqs.~(\ref{singlequbit}) and (\ref{nrhot}), one can immediately find
\begin{equation}
\lim_{t\rightarrow\infty}\rho_{\mathrm{GHZ}}(t)=\lim_{t\rightarrow\infty}\rho_{\mathrm{unc}}(t)=|g\rangle\langle g|^{\otimes N}.
\end{equation}
Due to the fact that this long-time steady state does not contain any information about the spectral density, we can conclude that $\mathcal{F}_{\theta}(\infty)=0$ and corresponding sensing error becomes divergent in the long-encoding-time regime. Such an error-divergence problem has been reported in previous works~\cite{PhysRevA.97.012126,PhysRevA.97.012125,Tamascelli_2020,SALARISEHDARAN2019126006,681442767}, which severely restricts the realization of precisely sensing the quantum reservoir.

\section{Threshold of the sensitivity}~\label{sec:sec5}
In the non-Markovian case, one needs to numerically solve Eq.~(\ref{ct}) to obtain the QFI. However, before performing these numerical simulations, we would like to use the Laplace transformation to analyze the long-time dynamical behavior of $c(t)$, which can provide some qualitative results and is benefit for us to establish a clear physical picture. Applying Laplace transformation on Eq.~(\ref{ct}), one can find $\tilde{c}(z)=[z+i\omega_{0}+\int_{0}^{\infty}\frac{J(\omega)}{z+i\omega}d\omega]^{-1}$. The expression of $c(t)$ can be derived by using the Cauchy residue theorem via finding the poles of $\tilde{c}(z)$ from
\begin{equation}~\label{transeq}
y(E)\equiv\omega_{0}-\int_{0}^{\infty}\frac{J(\omega)}{\omega-E}d\omega=E,~(E=iz).
\end{equation}
It can be proven that the roots of Eq. \eqref{transeq} are just the eigenenergies of the single sensor-reservoir Hamiltonian in the single-excitation subspace~\cite{PhysRevA.81.052330}. Therefore, the dynamical behavior of $c(t)$ is closely associated with the feature of the energy spectrum of the sensor-reservoir system. Since $y(E)$ is a monotonically decreasing function in the regime $E<0$, Eq.~(\ref{transeq}) has only one isolated root $E\equiv E_{b}$ provided $y(0)<0$. We call the eigenstate corresponding to this isolated eigenenergy the bound state. In contrast, Eq.~(\ref{transeq}) has infinite roots in the regime $E>0$, which form a continuous energy band. With the help of the above analysis, the inverse Laplace transform can be exactly done, which results in
\begin{equation}~\label{ctlaplace}
c(t)=Ze^{-iE_{b}t}+\int_{0}^{\infty}\frac{J(E)e^{-iE t}dE}{[E-\omega_{0}-\Delta(E)]^{2}+[\pi J(E)]^{2}},
\end{equation}
with $Z=[1+\int_{0}^{\infty}\frac{J(\omega)}{(E_{b}-\omega)^{2}}d\omega]^{-1}$. The first term in Eq.~(\ref{ctlaplace}) is contributed by the potentially formed bound state, while the second term is from the continuous band energies. The second term gradually vanishes with time due to out-of-phase interference. Thus, if a bound state is formed, then $\lim_{t\rightarrow\infty}c(t)=Ze^{-iE_{b}t}$, leading a dissipationless dynamics; while if the bound state is absent, then $\lim_{t\rightarrow\infty}c(t)=0$, which characterizes a complete decoherence. For the Ohmic-family spectral density considered in this paper, the criterion of forming a bound state can be analytically expressed as $\omega_{0}<\eta\omega_{c}\Gamma(s)$~\cite{PhysRevE.90.022122}, where $\Gamma(s)$ is Euler's $\Gamma$ function.

\begin{figure*}
\centering
\includegraphics[angle=0,width=0.95\textwidth]{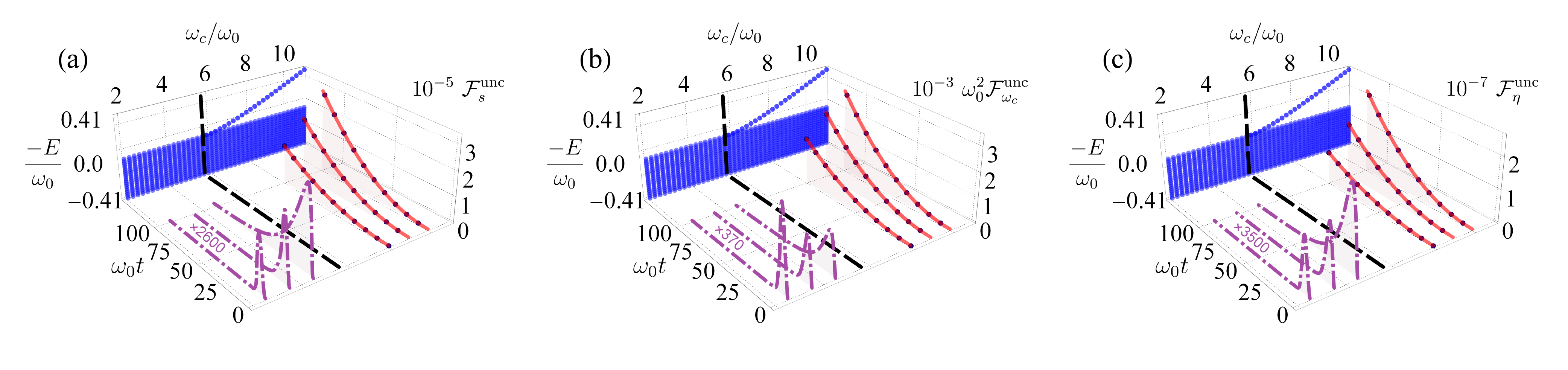}
\caption{Time evolution of $\mathcal{F}_{s}^{\mathrm{unc}}(t)$ in (a), $\mathcal{F}_{\omega_c}^{\mathrm{unc}}(t)$ in (b), and $\mathcal{F}_{\eta}^{\mathrm{unc}}(t)$ in (c) in different $\omega_c$. The blue dots are the energy spectrum of each TLS-reservoir system in single-excitation subspace. The black dashed line separates the regime into the areas with and without the bound state. The red solid and purple dot-dashed lines are the QFI by numerically solving Eq.~(\ref{ct}) when the bound state is present and absent, respectively. The purple dot-dashed lines are magnified by the times marked in the plots. The black circles represent the analytical results from Eq.~(\ref{uncfs}), which is in good agreement with the numerical results. We use $\eta=0.1$, $s=0.5$, and $N=100$.}\label{fig:fig1}
\end{figure*}

\begin{figure}
\centering
\includegraphics[angle=0,width=8.5cm]{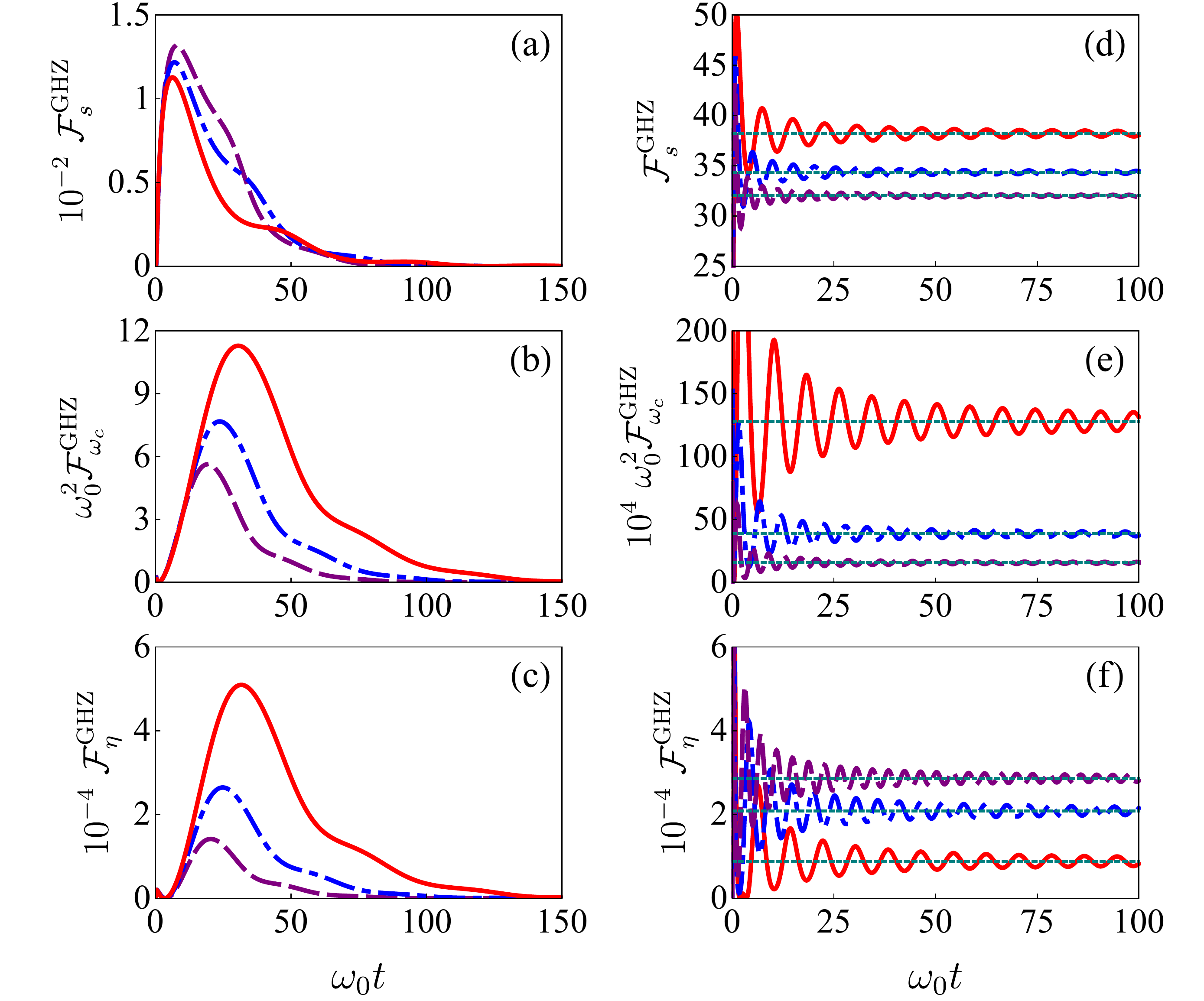}
\caption{\label{fig:fig2} Time evolution $\mathcal{F}_{s}^{\mathrm{GHZ}}(t)$ (a), $\mathcal{F}_{\omega_c}^{\mathrm{GHZ}}(t)$ (b), and $\mathcal{F}_{\eta}^{\mathrm{GHZ}}(t)$ (c) by numerically solving Eq.~(\ref{ct}) in the absence of the bound state when $\omega_{c}=7\omega_{0}$ (purple solid line), $7.5\omega_{0}$ (blue dot-dashed line), and $8\omega_{0}$ (red dashed line). The corresponding ones in (d), (e), (f) in the presence of the bound state when $\omega_{c}=20\omega_{0}$ (purple dashed line), $25\omega_{0}$ (blue dot-dashed line), and $\omega_{c}=30\omega_{0}$ (red solid line). The cyan dashed lines are analytical results from Eq.~(\ref{ghzfs}). We use $\eta=0.1$, $s=1$, and $N=200$.}
\end{figure}

Due to the fact that $\lim_{t\rightarrow\infty}c(t)=0$ in the absence of the bound state, our sensing scheme is then qualitatively consistent with the Markovian case, in which the sensing error diverges in the long-encoding-time regime irrespective of whether the initial state is the product state or the GHZ state. In the presence of the bound state, substituting the long-time form $\lim_{t\rightarrow\infty}c(t)=Ze^{-iE_{b}t}$ into Eqs. \eqref{uncff}, \eqref{ghzf1}, and \eqref{ghzf2}, we find that the long-time asymptotic QFI in the large-$N$ limit are given by (see Appendix \ref{dervat})
\begin{eqnarray}
\lim_{t\rightarrow\infty}\mathcal{F}_\theta^\text{unc}(t)&\simeq& NZ^{2}{E'_{b}}^{2}t^{2},\label{uncfs}\\
\lim_{t\rightarrow\infty}\mathcal{F}_{\theta}^{\mathrm{GHZ}}(t)&\simeq& \frac{2N{Z'}^{2}}{1-Z^{2}}.\label{ghzfs}
\end{eqnarray}
It is remarkable to see from the above expression that the QFI behaves as $\mathcal{F}_\theta^\text{unc}(t)\propto t^{2}$ when the initial state is the uncorrelated state. This result implies the sensing error can be continuously diminished by prolonging the encoding time, i.e., $\delta\theta\propto t^{-1}$. Such a time-scaling relation is the same as the noiseless Ramsey-spectroscopy metrology scheme~\cite{PhysRevLett.79.3865,Wang_2017}. It means that, much different from the previous work \cite{PhysRevA.97.012126,PhysRevA.97.012125,Tamascelli_2020,SALARISEHDARAN2019126006,681442767} and the Markovian approximate case, the encoding time in our scheme can be used as a resource to improve the sensing precision, which is similar to the widely used unitary-encoding scheme in the ideal case. On the other hand, if the initial state is the GHZ state, the QFI ultimately approaches a nonzero steady value. It is not a surprise that the product state outperforms the GHZ state in taking the encoding time as a resource because the entanglement contained in the GHZ state is fragile to the local decoherence. The resource of multipartite entanglement carried by the GHZ state has already been destroyed at the early evolution and cannot show its superiority in the long-time regime. Even so, both of the above results are completely different from the Markovian approximate result in which the QFI vanishes in the long-encoding-time regime. The bound state favored nonzero QFI in the long-time limit can be understood as follows. As a stationary state of the total system, the bound state is immune to the reservoir-induced decoherence. Therefore, the messages of spectral density encoded in the bound state is preserved and thus a nonzero QFI is achieved in the long-time limit. This sufficiently solves the error-divergency problem. Another result revealed by Eqs. \eqref{uncfs} and \eqref{ghzfs} is that the sensing precision scales with the number $N$ of the TLSs as the SNL in the large-$N$ limit. This result once again exhibits the advantage of our scheme over the Markovian approximate one, in which the SNL is totally destroyed in the long-time limit. It is noted that the distinguished roles played by the bound state in continuous-variable sensor \cite{PhysRevA.103.L010601} and the Mech-Zehnder-interferometry-based quantum metrology \cite{PhysRevLett.123.040402} have been reported.

To verify the above analytical results, we present the numerical calculations on the QFI as the function of $\omega_{0}t$ in Figs.~\ref{fig:fig1} and \ref{fig:fig2}. We can see that the QFI exhibits a similar behavior in the regime where the bound state is absent regardless of whether the initial state is the uncorrelated or GHZ states. At the beginning, the QFI gradually increases from its initial value $\mathcal{F}_{\theta}(0)=0$ to the maximal values. Then the QFI begins to decrease and eventually disappears in the long-encoding-time limit. This numerical result is consistent with many previous studies~\cite{PhysRevA.97.012126,PhysRevA.97.012125,Tamascelli_2020,SALARISEHDARAN2019126006,681442767}. In sharp contrast to this, when the bound state is present, the QFI for the uncorrelated state exhibits a square power-law increase with the encoding time (see Fig. \ref{fig:fig1}), while the one for the GHZ state saturates to a nonzero value [see Figs. \ref{fig:fig2}(d), \ref{fig:fig2}(e), and \ref{fig:fig2}(f)]. The numerical results in both Figs. \ref{fig:fig1} and \ref{fig:fig2} are in good agreement with our analytical Eqs. \eqref{uncfs} and \eqref{ghzfs} in the long-time limit, which validates our analytical results. It is noted that, although only the large-$N$ case is studied, the constructive role of the bound state in enhancing the sensing sensitivity is also valid in the small-$N$ case (see Appendix \ref{smancas}).

\section{Discussion and conclusions}~\label{sec:sec6}
It is necessary to point out that our sensing scheme is independent of the explicit form of the spectral density. Although only the Ohmic-family spectral densities are displayed in this paper, our sensing scheme can be generalized to other cases without difficulty. For non-Ohmic-family cases, the specific condition of forming a bound state might be quantitatively different, but the effectiveness of our scheme remains unchanged. Recently, the observation of the bound-state effect, which is the key recipe in our scheme, has been reported in circuit quantum electrodynamics architecture \cite{Liu2016} as well as matter-wave systems~\cite{Kri2018}. These experimental achievements provide a strong support to our sensing scheme and indicate that our finding is realizable by employing the experimental technique of quantum optics.

In summary, using $N$ TLSs as a sensor, we present a nonunitary-encoding sensing scheme to estimate the spectral density of a quantum reservoir. A mechanism to overcome the outstanding error-divergence problem in the literature is revealed. It is found that, accompanying the formation of a TLS-reservoir bound state, the shot-noise-limited estimation error saturates to a finite value for the GHZ-type entangled state and persistently decreases for the product state with the encoding time. An analytical threshold to achieve this amazing result is given explicitly. Our result supplies an insightful guideline to practically realize the sensing to the quantum reservoir and may have a deep impact on controlling the reservoir-induced decoherence to microscopic quantum systems.

\section*{Acknowledgments}
The work is supported by the National Natural Science Foundation (Grants No. 11704025, No. 11875150, No. 11834005, and No. 12047501).

\appendix

\begin{widetext}
\section{Derivation of QFI}\label{dervat}
We give the detailed derivation of the QFI here. In the uncorrelated state case, one can find that the reduced density matrix of the sensor is an $N$-fold tensor product of $\rho_{\mathrm{single}}(t)$, i.e., $\rho_{\mathrm{unc}}(t)=\rho_{\mathrm{single}}(t)^{\otimes N}$ with
\setcounter{equation}{0}
\renewcommand\theequation{A\arabic{equation}}
\begin{equation}~\label{app:app1}
\begin{split}
\rho_{\mathrm{single}}(t)=&\frac{p_{t}}{2}|e\rangle\langle e|+\Big{(}1-\frac{p_{t}}{2}\Big{)}|g\rangle\langle g|+\bigg[\frac{c(t)}{2}|e\rangle\langle g|+\mathrm{H}.\mathrm{c}.\bigg],
\end{split}
\end{equation}
and $p_t=|c(t)|^2$. According to the additivity feature of QFI for product state, we have $\mathcal{F}_{\theta}^{\mathrm{unc}}(t)=N\mathcal{F}_{\theta}^{\mathrm{single}}(t)$. Thus the derivation of $\mathcal{F}_{\theta}^{\mathrm{unc}}(t)$ is simplified to calculate the QFI of a single-TLS state $\rho_{\mathrm{single}}(t)$. One can see that $\rho_{\mathrm{single}}(t)$ is a two-by-two matrix in the basis $\{|e\rangle,|g\rangle\}$. It has been found that the QFI for such state $\rho_{\mathrm{single}}(t)$ with the dimension of Hilbert space being two relates to the Bloch vector $\mathbf{r}(t)=\big{(}\mathrm{Re}[c(t)],-\mathrm{Im}[c(t)],p_{t}-1\big{)}^{\mathrm{T}}$ as~\cite{Liu_2019}
\begin{equation}~\label{app:app2}
\mathcal{F}_{\theta}^{\mathrm{single}}(t)=|\mathbf{r}'(t)|^2+\frac{[\mathbf{r}(t)\cdot{\mathbf{r}'(t)}]^2 }{1-|\mathbf{r}(t)|^2}.
\end{equation}
When the bound state is formed, the substitution of $c(\infty)\simeq Ze^{-iE_{b} t}$ into Eq. \eqref{app:app2} results in
\begin{equation}~\label{app:app3}
 \mathcal{F}_{\theta}^{\mathrm{single}}(\infty)\simeq \frac{2Z(Z^{2}-2)Z'}{Z^{2}-1}+Z^{2}{E'_{b}}^{2}t^{2}.
\end{equation}
The first term on the right side of Eq. \eqref{app:app3} is independent of time and becomes much smaller than the second term in the long-time regime. Therefore, we can safely drop the first term and finally obtain
\begin{equation}~\label{app:app4}
\lim_{N\rightarrow\infty}\mathcal{F}_{\theta}^{\mathrm{unc}}(\infty)\simeq NZ^{2}{E'_{b}}^{2}t^{2}.
\end{equation}

When the initial state is the GHZ state, we find the reduced density matrix of the sensor is given by
\begin{equation}~\label{app:app5}
  \rho_{\mathrm{GHZ}}(t)=\frac{1}{2}\Big{[}p_{t}|e\rangle\langle e|+(1-p_{t})|g\rangle\langle g|\Big{]}^{\otimes N}+\frac{1}{2}\Big{[}|g\rangle\langle g|^{\otimes N}+\Big{(}c(t)^{N}|e\rangle\langle g|^{\otimes N}+\mathrm{H}.\mathrm{c}.\Big{)}\Big{]}.
\end{equation}
Expanding the first term of Eq. \eqref{app:app5} as~\cite{PhysRevA.84.022302,PhysRevA.87.032102}
\begin{equation}~\label{app:app6}
\begin{split}
\big{[}p_{t}|e\rangle\langle e|+(1-p_{t})|g\rangle\langle g|\big{]}^{\otimes N}=&p_{t}^{N}|e\rangle\langle e|^{\otimes N}+\big{(}1-p_{t}\big{)}^{N}|g\rangle\langle g|^{\otimes N}\\
&+\sum_{\mathbb{P}}\sum_{m=1}^{N-1}p_{t}^{m}\big{(}1-p_{t}\big{)}^{N-m}
\mathbb{P}\big{[}|e\rangle\langle e|^{\otimes m}\otimes|g\rangle\langle g|^{\otimes(N-m)}\big{]},
\end{split}
\end{equation}
where $\mathbb{P}$ represents all possible bipartite permutations, we can rearrange Eq. \eqref{app:app5} as $\rho_{\mathrm{GHZ}}(t)=\rho_{1}(t)\oplus\rho_{2}(t)$ with
\begin{equation}~\label{app:app7}
\rho_{1}(t)=\frac{1}{2}\big{\{}p_{t}^{N}|e\rangle\langle e|^{\otimes N}+\big{[}1+(1-p_{t})^{N}\big{]}|g\rangle\langle g|^{\otimes N}+\big{[}c(t)^{N}|e\rangle\langle g|^{\otimes N}+\mathrm{H}.\mathrm{c}.\big{]}\big{\}},
\end{equation}
and
\begin{equation}~\label{app:app8}
\rho_{2}(t)=\frac{1}{2}\sum_{\mathbb{P}}\sum_{m=1}^{N-1}p_{t}^{m}\big{(}1-p_{t}\big{)}^{N-m}\mathbb{P}\big{[}|e\rangle\langle e|^{\otimes m}\otimes|g\rangle\langle g|^{\otimes(N-m)}\big{]}.
\end{equation}
It is easy to see that $\rho_{1}(t)$ is a two-by-two matrix in the basis $\{|e\rangle^{\otimes N},|g\rangle^{\otimes N}\}$ and $\rho_{2}(t)$ is a $(2^{N}-2)$-dimensional diagonal matrix. According to the fact that the QFI is additive for a direct-sum density matrix, we have $\mathcal{F}_{\theta}^{\mathrm{GHZ}}(t)=\mathcal{F}_{\theta}^{(1)}+\mathcal{F}_{\theta}^{(2)}$ with $\mathcal{F}_{\theta}^{(1,2)}$ being the QFI of $\rho_{1,2}(t)$.

The QFI for an arbitrary state $\rho$ reads \cite{Liu_2019}
\begin{equation}~\label{app:app9}
\mathcal{F}_{\theta}=\sum_{i,j=1}^{M}\bigg{[}\frac{{\lambda'}_{i}^{2}}{\lambda_{i}}+4\lambda_{i}\langle\lambda'_{i}|\lambda'_{i}\rangle-\frac{8\lambda_{i}\lambda_{j}}{\lambda_{i}+\lambda_{j}}|\langle\lambda_{i}|\lambda'_{j}\rangle|^{2}\bigg{]},
\end{equation}where $\rho=\sum_{i=1}^{M}\lambda_{i}|\lambda_{i}\rangle\langle\lambda_{i}|$. It can be proven that Eq. \eqref{app:app9} returns to Eq. \eqref{app:app2} in the one-TLS case. With Eq. \eqref{app:app9} and the eigendecomposition of $\rho_{1}(t)$ at hand, $\mathcal{F}_{\theta}^{(1)}$ is readily calculated. Substituting $c(\infty)\simeq Ze^{-iE_{b} t}$ in the presence of the bound state into Eq. \eqref{app:app7}, one can find $\rho_{1}(\infty)$ reduces to
\begin{equation}~\label{app:app11}
\rho_{1}(\infty)\simeq \frac{1}{2}\left[
                                    \begin{array}{cc}
                                      Z^{2N} & Z^{N}e^{-iNE_{b}t} \\
                                      Z^{N}e^{iNE_{b}t} & 1+(1-Z^{2})^{N} \\
                                    \end{array}
                                  \right].
\end{equation}
Although Eq. \eqref{app:app11} can contribute a time-dependent QFI, it tends to
\begin{equation}~\label{app:app12}
\lim_{N\rightarrow\infty}\rho_{1}(\infty)\simeq \frac{1}{2}\left(
                                    \begin{array}{cc}
                                      0 & 0 \\
                                      0 & 1 \\
                                    \end{array}
                                  \right),
\end{equation}
in the large-$N$ limit due to the fact $\lim_{N\rightarrow\infty}Z^{N}=0$. Equation \eqref{app:app12} is independent of $\theta$ and thus has no contribution to the QFI, i.e., $\lim_{N\rightarrow\infty}\mathcal{F}_{\theta}^{(1)}(\infty)=0$. On the other hand, since the eigenvectors of the diagonal matrix $\rho_{2}(t)$ are independent of $\theta$, the nonzero component is only the first term of Eq. \eqref{app:app9} as
\begin{eqnarray}~\label{app:app10}
  \mathcal{F}_{\theta}^{(2)}&=&\frac{1}{2}\sum_{m=1}^{N-1}\frac{N!}{m!(N-m)!}\frac{\{[p_{t}^{m}(1-p_{t})^{N-m}]'\}^{2}}{p_{t}^{m}(1-p_{t})^{N-m}}=\frac{p_{t}^{\prime2}}{2}\sum_{m=1}^{N-1}\frac{N!}{m!(N-m)!}(1-p_{t})^{N-m-2}p_{t}^{m-2}(m-Np_{t})^{2}\nonumber\\
  &=&-\frac{Np_{t}^{\prime2}}{2 p_{t}}\frac{}{}\bigg{\{}\frac{[Np_{t}(1-p_{t})^{N}+p_{t}-1]}{(p_{t}-1)^{2}}+Np_{t}^{N-1}\bigg{\}}.
\end{eqnarray}
The substitution of $c(\infty)\simeq Ze^{-iE_{b} t}$ into Eq.~(\ref{app:app10}) results in
\begin{equation}~\label{app:app13}
  \mathcal{F}_{\theta}^{(2)}(\infty)\simeq2Z^{2}{Z'}^{2}\Big{[}\frac{N}{Z^2-Z^4}-N^{2}Z^{2N-4}-N^{2}(1-Z^2)^{N-2}\Big{]},
\end{equation}
which is independent of the encoding time regardless of whether $N$ is small or large. In the large-$N$ limit, Eq.~(\ref{app:app13}) can be further simplified to
\begin{equation}~\label{app:app14}
\lim_{N\rightarrow\infty}\mathcal{F}_{\theta}^{(2)}(\infty)\simeq\frac{2N{Z'}^{2}}{1-Z^{2}}.
\end{equation}

\begin{figure*}
\centering
\includegraphics[angle=0,width=0.96\textwidth]{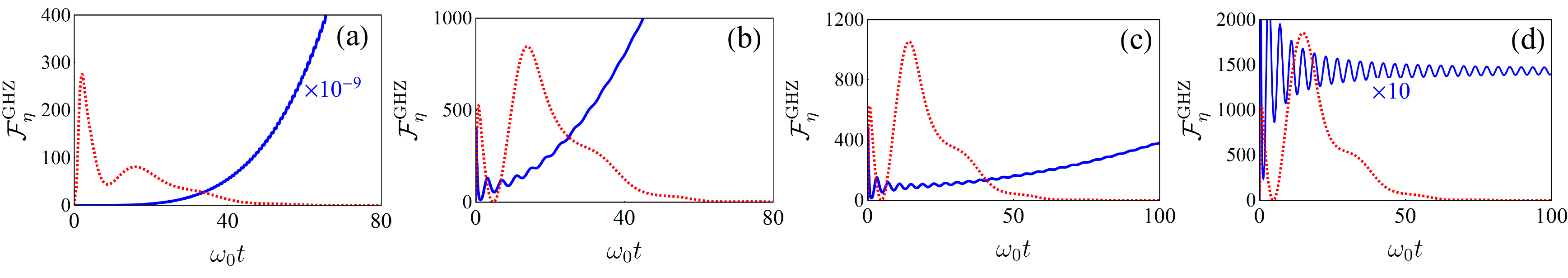}
\caption{ Time evolution of $\mathcal{F}_{\eta}^{\mathrm{GHZ}}(t)$ when $N=5$ (a), $50$ (b), $60$ (c), and $100$ (d). The blue and red dotted lines are the results with the bound state when $\omega_{c}=30\omega_{0}$ and without the bound state when $\omega_{c}=6\omega_{0}$, respectively. The blue solid lines are magnified by the times marked in the plots. We use $\eta=0.1$ and $s=1$.}\label{fig:fig3}
\end{figure*}
\section{QFI in the small-$N$ regime}\label{smancas}

In the main text, we show the bound-state-favored quantum sensing of the reservoir in the large-$N$ limit. It is noted that the constructive role of the bound state in enhancing the sensing sensitivity is also present in the small-$N$ limit. Taking $\mathcal{F}_{\eta}^{\mathrm{GHZ}}(t)$ as an example, we plot in Fig. \ref{fig:fig3} its evolution in different $N$. In the absence of the bound state, $\mathcal{F}_{\eta}^{\mathrm{GHZ}}(t)$ tends to zero irrespective of the value of $N$. In contrast, $\mathcal{F}_{\eta}^{\mathrm{GHZ}}(t)$ becomes larger and larger with time in the small $N$ case when the bound state is formed. As analyzed in the last section, such persistent increasing QFI is contributed from $\rho_{1}(\infty)$ in Eq. \eqref{app:app11}. With the increasing of $N$, the QFI from $\rho_{1}(\infty)$ becomes smaller and smaller and tends to zero in the large-$N$ limit. Therefore, $\mathcal{F}_{\eta}^{\mathrm{GHZ}}(\infty)$ in this limit contains only the contribution from $\rho_{2}(\infty)$, which is a constant [see Eq. \eqref{app:app14}]. The numerical results in Fig.~\ref{fig:fig3} are in good agreement with our above analytical solution, which validates our conclusion.

\end{widetext}
	
\bibliography{reference}

\end{document}